\documentclass[%
 reprint,
nofootinbib,
 amsmath,amssymb,
 aps,
floatfix,
]{revtex4-2}

\usepackage{graphicx}
\usepackage{xcolor}
\usepackage{dcolumn}
\usepackage{bm}
\usepackage{upgreek}
\usepackage{physics}
\usepackage[caption=false]{subfig}
\usepackage{hyperref}
\hypersetup{colorlinks,linkcolor={blue},citecolor={blue},urlcolor={blue}}
\usepackage{xurl}
\usepackage[mathlines]{lineno}
 
\begin{document}

\preprint{NITEP 230}

\title{Investigating nuclear density profiles to reveal particle-hole configurations \\in the island of inversion}%

\author{R. Barman$^{1,2}$, W. Horiuchi$^{3,4,5,2}$, M. Kimura$^2$, and R. Chatterjee$^1$}
\affiliation{
  $^1$Department of Physics, Indian Institute of Technology Roorkee, Roorkee 247 667 India\\
  $^2$RIKEN Nishina Center, Wako 351-0198, Japan\\
  $^3$Department of Physics, Osaka Metropolitan University, Osaka 558-8585, Japan\\
$^4$Nambu Yoichiro Institute of Theoretical and Experimental Physics (NITEP), Osaka Metropolitan University, Osaka 558-8585, Japan\\
  $^5$Department of Physics, Hokkaido University, Sapporo 060-0810, Japan}

\date{\today}

\begin{abstract}
\noindent
\textbf{Background:}
In the mass regions with an abnormal shell structure, the so-called ``island of inversion," the spin-parity of odd-mass nuclei provides quantitative insights into the shell evolution. However, the experimental determination of the spin-parity is often challenging, leaving it undetermined in many nuclei. 

\textbf{Purpose:}
We discuss how the shell structure affects the density profiles of nuclei in the island of inversion and investigate whether these can be probed from the total reaction and elastic scattering cross sections.

\textbf{Method:}
The antisymmetrized molecular dynamics (AMD) is employed to generate various particle-hole configurations and predict the energy levels of these nuclei.
The obtained density distributions are used as inputs to the Glauber model, which is employed to calculate the total reaction and elastic scattering cross sections for revealing their relationship to the particle-hole configurations.
   
\textbf{Results:}
In addition to the well-known correlation between nuclear deformation and radius, we show the correlations between the particle-hole configurations and both central density and diffuseness.
We show that different particle-hole configurations are well reflected in the total reaction and elastic scattering cross sections.

\textbf{Conclusion:} The total reaction and elastic scattering cross sections are useful probes to identify the spin-parity of nuclei when different particle-hole configurations coexist.
 
\end{abstract}

\maketitle

\section{\label{sec:intro}INTRODUCTION}
Recent experimental advances have revealed an unusual structure evolution of exotic nuclei through investigations of nuclear density distributions. Various exotic features, including the halo~\cite{tanihata_halo_1985, Hansen_1987, IsaoTanihata_1996, Tanaka_PhysRevLett.104.062701, TOGANO2016412, bagchi_2020, nakamura_PhysRevLett.103.262501} and skin~\cite{suzuki_halo_1995, kanungo_PhysRevLett.117.102501, BAGCHI2019251, Kaur_PhysRevLett.129.142502, tanaka_frontiers} structure, have been identified by systematic measurements of nuclear radii using the total reaction cross sections. These cross sections are also sensitive to nuclear deformation, and the evolution of nuclear deformation in various neutron-rich nuclei has been extensively discussed through total reaction cross section measurements~\cite{TAKECHI2012357, horiuchi_PhysRevC.81.024606, Takechi_2014, minomo_PhysRevC.84.034602, minomo_PhysRevLett.108.052503, sumi_PhysRevC.85.064613,  horiuchi_PhysRevC.86.024614, watanabe_PhysRevC.89.044610, horiuchi_PhysRevC.105.014316, takatsu_PhysRevC.107.024314}.

The nuclear diffuseness parameter is another crucial quantity that characterizes the nuclear density distribution near the surface and provides valuable information on the nuclear structure. Since the diffuseness influences the height of the first peak in the elastic scattering cross sections~\cite{Amado_PhysRevC.21.647, hatakeyama_PhysRevC.97.054607}, its measurement offers a potential probe for the nuclear structure. For example, Ref.~\cite{vishal_bubble_2020} discussed the possibility of determining the nuclear bubble structure by estimating the internal density distributions from the nuclear diffuseness. Furthermore, Ref.~\cite{vc_2021} demonstrated that the diffuseness parameter varies with particle-hole configurations in the isotopic chains of Ne and Mg. 

Motivated by these findings, we explore the possibility of identifying particle-hole configurations of unknown nuclei, particularly in the island of inversion, through the analysis of the total reaction and elastic scattering cross sections. The idea of the island of inversion was first proposed~\cite{ioi_PhysRevC.41.1147} to explain the anomalous mass behavior and the low-lying first $2^+$ states in the even-even nuclei around $^{32}$Mg~\cite{thibault_PhysRevC.12.644, CAMPI1975193, detraz_PhysRevC.19.164}. Subsequent studies further highlighted the onset of nuclear deformation in this region through the measurement of large electric-quadrupole transition strengths associated with the first $2^+$ state~\cite{MOTOBAYASHI19959, IWASAKI2001227, YANAGISAWA200384}. In this region, the intruder configurations dominate the ground state, resulting in the coexistence of many-particle many-hole ($m$p$n$h) configurations~\cite{kimura-horiuchi_2004, kimura_PhysRevC.75.041302, neyens_PhysRevC.84.064310, nishibata_2019, KITAMURA2021136682}.   
In such a situation, the spectra of odd-mass nuclei provide crucial information on the shell evolution. However, the experimental determination of the spin-parity remains challenging even for the ground states.

In this paper, we examine whether the particle-hole configurations of the ground states can be identified from the total reaction and elastic scattering cross sections. We employ the antisymmetrized molecular dynamics (AMD) to obtain various particle-hole configurations and their density distributions. Using these density distributions, we calculate the total reaction and elastic scattering cross sections within the framework of the Glauber model. We first verify the feasibility of our approach in $^{31}$Mg as a test case, where the particle-hole configurations of the ground- and low-lying excited states are well known. We then extend our discussion to $^{29}$Ne, $^{33}$Mg, and $^{35}$Mg, where the ground-state particle-hole configurations and the spin-parity assignments are uncertain.

This paper is organized as follows. In the next section, we briefly 
describe AMD
 and the Glauber model. In Section \ref{subsection:IIIA}, we validate the correlation between cross sections and particle-hole configurations of $^{31}$Mg as a test case. Subsequently, in Sec.~\ref{subsection:IIIB}, we discuss $^{29}$Ne, $^{33}$Mg, and $^{35}$Mg, where the spin-parity assignments are not well established. Finally, we summarize this work in Sec.~\ref{sec:conclusion}.

 \section{FORMALISM}\label{sec:formalism}
 \subsection{\label{sec:AMD}AMD + GCM framework}
 In this study, we employ AMD~\cite{Kimura2016, enyo_AMD_2012} to calculate various particle-hole configurations and their density distributions. The Hamiltonian for the AMD calculation is defined as
 \begin{align}
   \hat{H} = \sum_{i=1}^{A} \hat{t}_i - \hat{t}_{\rm cm} + \sum_{i<j}^{A} \hat{v}^{N}_{ij} + \sum_{i<j}^{A} \hat{v}^{C}_{ij}.
 \end{align}
 Here, $\hat{t}_i$ is the single-particle kinetic energy, $\hat{v}^N_{ij}$ is an effective nucleon-nucleon interaction with Gogny D1S parameterization, and $\hat{v}^C_{ij}$ is the Coulomb interaction. The center-of-mass kinetic energy $\hat{t}_{\rm cm}$ is exactly removed.
 The intrinsic AMD wave function is represented by a Slater determinant of nucleon wave packets
 \begin{align}
   \Phi_{\rm int} &= \mathcal{A} \{\varphi_1, \varphi_2,..., \varphi_A\}, \\
   \varphi_i(\bm r) &= \phi_i(\bm r)\chi_i \xi_i.
 \end{align}
 The wave packet $\varphi_i$ consists of the spatial part ($\phi_i$) with a Gaussian form; and spin ($\chi_i$) and isospin ($\xi_i$) parts
 \begin{align}
 \phi_i(\bm r)&=\exp\left\{-\sum_{\sigma=x,y,z}\nu_{\sigma}\left(r_{\sigma}-\frac{Z_{i\sigma}}{\sqrt{\nu_{\sigma}}}\right)^2\right\},\\
 \chi_i &= a_i \chi_{\uparrow} + b_i \chi_{\downarrow},\\
 \xi_i &= \mathrm{proton\ or\ neutron}, 
 \end{align}
 where the centroids of the Gaussian $Z_{i\sigma}$, the spin direction $a_i$ and
 $b_i$; and the width parameters $\nu_x$, $\nu_y$, $\nu_z$ are the variational parameters. The variational wave function is the parity-projected wave function,  $\Phi^{\pi}=\hat{P}^{\pi}\Phi_{\rm int}$, where $\hat{P}^{\pi}\ (\pi=\pm)$ is the parity projector. 
 
 The variational parameters are determined by the energy variation with the constraint on the matter quadrupole deformation parameter $\beta$~\cite{dote_PhysRevC.56.1844, Suhara_2010, kimura_10.1143/PTP.127.287}. We do not put any constraint over the quadrupole deformation parameter $\gamma$, and thus, it has an optimal value for each $\beta$. We obtain the optimized wave function $\Phi^{\pi}(\beta)$, which has the minimum energy and given value of the deformation parameter $\beta$. The detailed explanation of $\beta$ and $\gamma$ are also given in the Appendix A.
 After the variational calculation, $\Phi^{\pi}(\beta)$ are projected to the eigenstate of the angular momentum,
 \begin{align}
 \Phi_{MK}^{J\pi}(\beta) ={\hat{P}}_{MK}^J \Phi^{\pi}(\beta),
 \end{align}
 where ${\hat{P}}_{MK}^J$ is the angular momentum projector. Then, we superpose $\Phi_{MK}^{J\pi}(\beta)$
 \begin{align}\label{eq:8}
 \Psi_{Mn}^{J\pi} = \sum_{iK}C_{iKn} \Phi_{MK}^{J\pi}(\beta_i).
 \end{align}
 The coefficients $C_{iKn}$ and the eigen-energies are determined by solving Hill-Wheeler equation \cite{hw1953}
 \begin{align}
   \sum_{i'K'} H_{iKi'K'}C_{i'K'n} = E_n \sum_{i'K'} N_{iKi'K'}C_{i'K'n},
 \end{align}
 where the matrix elements of the Hamiltonian and norm are given by
 \begin{align}
   H_{iKi'K'}&= \langle\Phi^{J\pi}_{MK}(\beta_{i})|\hat{H}|\Phi^{J\pi}_{MK'}(\beta_{i'})\rangle,\\
   N_{iKi'K'}&= \langle\Phi^{J\pi}_{MK}(\beta_{i})|\Phi^{J\pi}_{MK'}(\beta_{i'})\rangle.
 \end{align}

 The nucleon density distributions are obtained from the following expression: \begin{align} \rho_{JM} (\bm r) &=  \left<\Psi_{Mn}^{J\pi}\left|\delta^{3}(\bm r - \bm r_{\rm cm} - \bm r_{i})\right|\Psi_{Mn}^{J\pi}\right>,\label{eq:12}\\
  &= \sum_{\lambda}C^{JM}_{\lambda 0 JM}\rho^{\lambda}_{J}(r)Y_{\lambda 0}^*(\hat{r}), \label{eq:13}
 \end{align}
where $\bm r_{\rm cm}$ is the coordinate of the center-of-mass and $r_i$ is the coordinate of the nucleon from the origin. Although there can be $\lambda>0$ (non-spherical) components, the $\lambda=0$ component in Eq.~(\ref{eq:13}) is used as an input to the Glauber model.

 \subsection{Glauber model}

The total reaction and elastic scattering cross sections are evaluated by the Glauber model~\cite{glauber1959}.
The total reaction cross section is expressed by
\begin{align}
\sigma_R=\int d\bm{b} \left(1-|e^{i\chi(\bm{b})}|^2\right),
\end{align}
where $\bm{b}$ is the two-dimensional impact parameter vector perpendicular to the beam direction, $z$,
and $e^{i\chi(\bm{b})}$ is the optical phase-shift function, which includes all dynamical information
within the adiabatic and eikonal approximations made in the Glauber theory.
With the help of the optical limit approximation, the phase-shift function casts into the following simple form
for proton scattering:
\begin{align}
e^{i\chi(\bm{b})}&=\exp\left[-\int d\bm{r} \left[\rho_{\rm n}(\bm{r})\Gamma_{\rm pn}(\bm{b}+\bm{s}) + \rho_{\rm p}(\bm{r})\right. \right.\nonumber\\
& \left. \left. \vphantom{\rho_{\rm n}(\bm{r})} \times\Gamma_{\rm pp}(\bm{b}+\bm{s})\right] \vphantom{\int}\right],
\end{align}
where $\bm{r}=(\bm{s},z)$ and $\Gamma_{NN}$ is the profile function, which describes the nucleon-nucleon scattering properties.
The standard parameter set is tabulated in Ref.~\cite{ibrahim_2008} for proton-neutron ($NN={\rm pn})$ and proton-proton 
$(NN={\rm pp}$), and $\Gamma_{\rm pn}=\Gamma_{\rm np}$ and $\Gamma_{\rm pp}=\Gamma_{\rm nn}$.
The angular distribution of the elastic scattering cross section can also be calculated by using
the phase-shift function as
\begin{align}
\frac{d\sigma}{d\Omega}(\theta)=|F(\theta)|^2
\end{align}
with the scattering amplitude~\cite{glauber1959,suzuki03}
\begin{align}
F(\theta)=F_C(\theta)+\frac{iK}{2\pi}\int d\bm{b}\,e^{-i\bm{q}\cdot\bm{b}+i\chi_C(\bm{b})}(1-e^{i\chi(\bm{b})}),
\end{align}
where $\bm{q}$ is the momentum transfer vector with $q=2K\sin(\theta/2)$, $F_C$ is the Rutherford scattering amplitude, and $i\chi_C$ is the Coulomb phase of the colliding two charged particles.
For a carbon target,
we use the nucleon-target profile function in the Glauber theory~\cite{horiuchi_PhysRevC.75.044607} for the phase-shift function, like
\begin{align}
e^{i\chi(\bm{b})} &\approx \exp\left\{-\int d\bm{r} \rho(\bm{r}) \left[ 1 - \exp\left(-\int d\bm{r}^\prime \right. \right. \right. \nonumber \\
& \left. \left. \left. \vphantom{\int} \times\rho_T(\bm{r}^\prime) \Gamma_{NN}(\bm{b} + \bm{s} - \bm{s}^\prime) \right) \right] \right\}
\end{align}
where $\rho_{T}$ is the density distribution of the target nucleus.
Note that the expression is simplified by omitting the sum for proton and neutron, 
and the symmetrized expression for the projectile and target nuclei is used. 
For the complete expression, the reader is referred to Ref.~\cite{ibrahim_2000}.
This expression gives a good description of medium- to high-energy nucleus-nucleus reactions~\cite{horiuchi_PhysRevC.75.044607, ibrahim_2009, horiuchi_PhysRevC.86.024614, nagahisa_PhysRevC.97.054614, horiuchi_PhysRevC.105.014316},
while the inputs to this expression are the same as the conventional optical approximation.
In the present work, we use a $^{12}$C target of the harmonic-oscillator type density distribution that is
set to reproduce the measured charge radius.
It should be noted that the theory has no adjustable parameter after all the inputs are set
and has succeeded in describing medium- to high-energy nuclear reactions involving unstable nuclei~\cite{horiuchi_PhysRevC.86.024614, horiuchi_aris2014, horiuchi_PhysRevC.93.044611, horiuchi_PhysRevC.96.024605, nagahisa_PhysRevC.97.054614, horiuchi_PhysRevC.74.034311, horiuchi_PhysRevC.75.044607, ibrahim_2009, horiuchi_PhysRevC.81.024606}.
The phase-shift function only requires the density distributions of the projectile nucleus,
and thus the computed cross sections crucially reflect the properties of the density profiles.

\section{Results and discussion}\label{sec:results}

\subsection{Particle-hole configurations of $^{31}\rm Mg$ and their impact on the cross sections}\label{subsection:IIIA}

Firstly, we review the properties of particle-hole configurations of $^{31}\rm Mg$ calculated by AMD, which are consistent with those previously obtained results~\cite{kimura_PhysRevC.75.041302}. Then, we discuss how the particle-hole configurations affect the density profile and cross sections, which is a new analysis presented in this work. 

\begin{figure*}[tb]
    \begin{center}
    \includegraphics[width=0.8\textwidth]{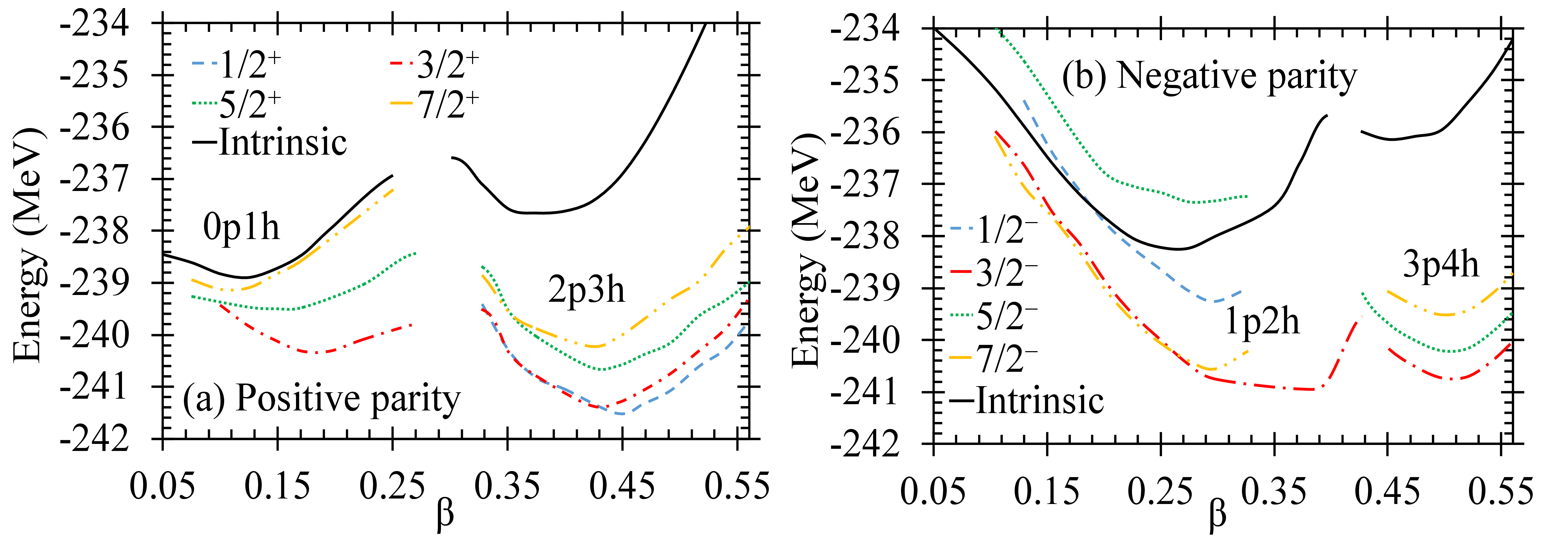}
    \caption{Binding energy curves of $^{31}$Mg for (a) the positive- and (b) negative-parity states. The solid curves show the energies before the angular momentum projection, while the broken lines show those after the angular momentum projection}
    \label{fig:Mg31surf}
    \end{center}
\end{figure*}
\begin{figure*}[tb]
    \begin{center}
    \includegraphics[width=0.8\textwidth]{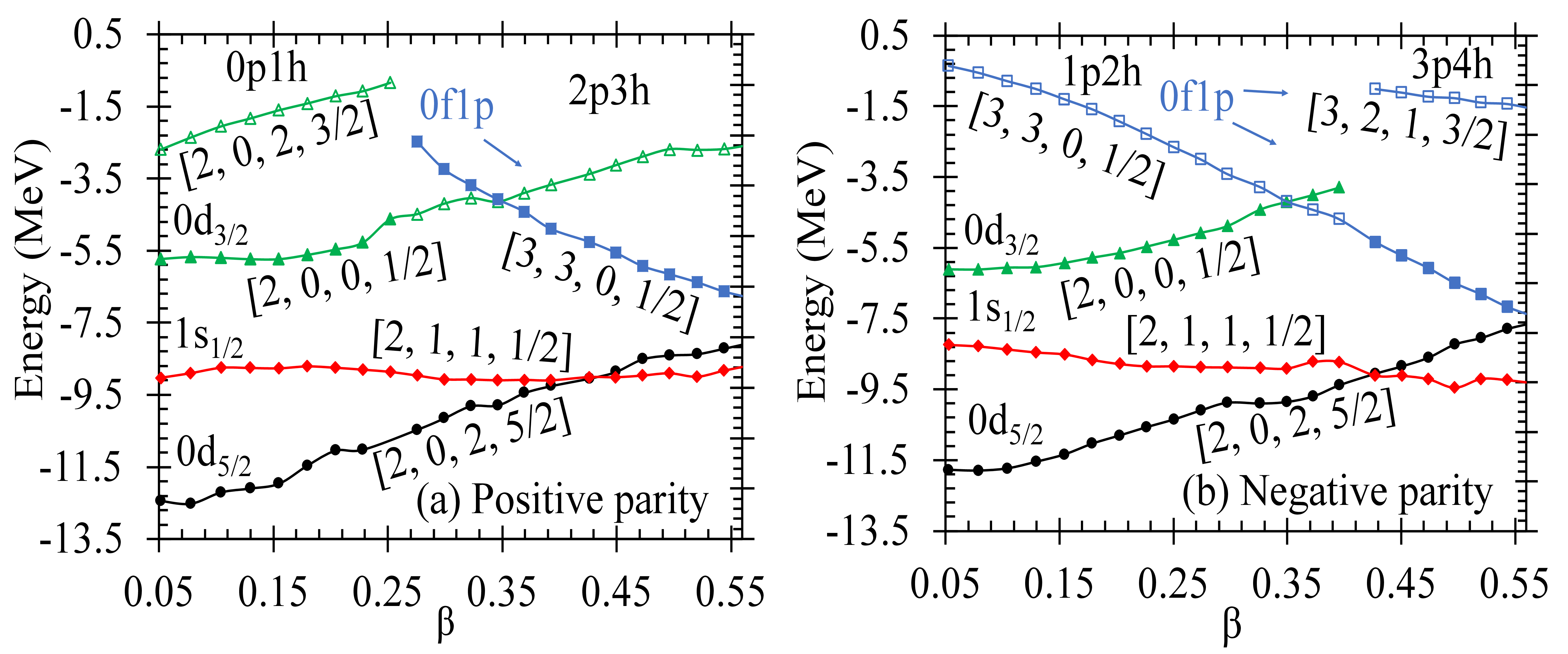}
    \caption{Neutron single particle levels of the last 7 neutrons of $^{31}$Mg. The open (filled) symbols represent the orbits occupied by one (two) neutron(s).}
    \label{fig:Mg31spo}
    \end{center}
\end{figure*}
Figure~\ref{fig:Mg31surf} shows the binding energy curves for the positive- and negative-parity states of $^{31}$Mg
as a function of the quadrupole deformation parameter $\beta$. 
In both parity states, different particle-hole configurations coexist. They are labeled for each energy minima and can be understood from the neutron single-particle orbits shown in Fig.~\ref{fig:Mg31spo}. Since the behaviors of the single-particle orbits can be qualitatively understood by the Nilsson model~\cite{bohr-mottelson}, it is helpful to classify them using the asymptotic quantum numbers $[n, n_z, l_z, \Omega]$.
The positive-parity states with $\beta \leq 0.25$ have a hole in the orbit with $[n, n_z, l_z, \Omega]=$ $[2,0,2,3/2]$. Hence, they correspond to 0p1h configurations. 
The largely deformed states with $\beta > 0.25$ have an intruder orbit with $[3,3,0,1/2]$ occupied by two neutrons, giving a 2p3h configuration. Since the 0p1h configuration has the last neutron in the [2, 0, 2, 3/2] orbit, it produces a $J^\pi=3/2^{+}$ state followed by $5/2^{+}$ and $7/2^{+}$ states, while 2p3h configuration generates a $K^{\pi}=1/2^+$ rotational band. We can also see that the negative-parity states have 1p2h ($\beta\leq0.4$) and 3p4h ($\beta > 0.4$) configurations in ascending order of deformation, which generate a group of the $3/2^-$, $7/2^-$, $1/2^-$, and $5/2^-$ states and a $K^\pi= 3/2^-$ rotational band, respectively. Among all these configurations, the $1/2^{+}$ state with 2p3h configuration is the lowest energy state.

\begin{figure}[htb]
    \begin{center}
    \includegraphics[width=\hsize]{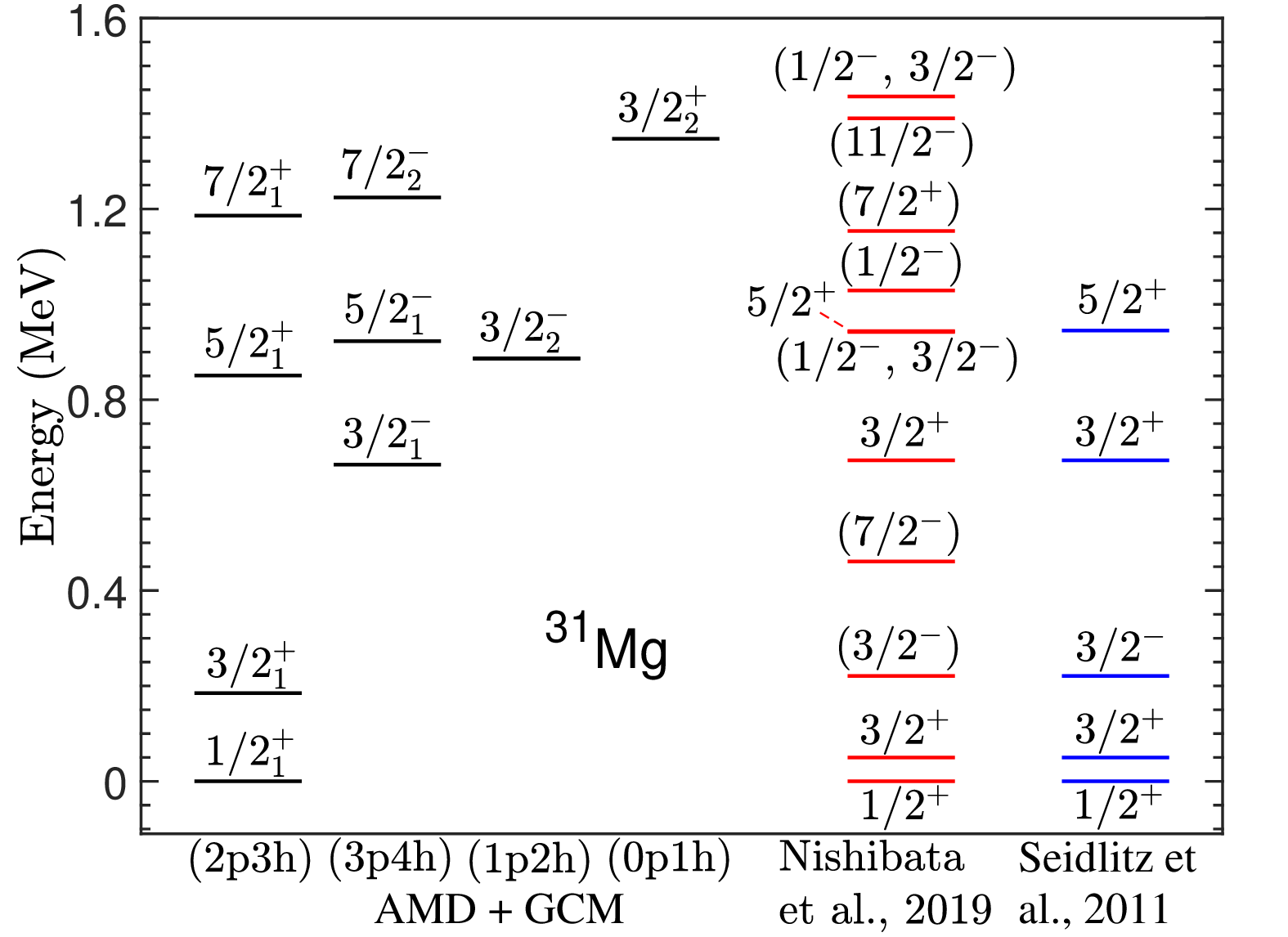}
    \caption{Calculated and observed energy spectra of $^{31}\mathrm{Mg}$. Experimental data are taken from Refs.~\cite{nishibata_2019, KITAMURA2021136682}.}
    \label{fig:ex31Mg}
    \end{center}
\end{figure}

Figure \ref{fig:ex31Mg} shows the energy spectrum obtained by the GCM calculations.
These results agree with those obtained in the previous work~\cite{kimura_PhysRevC.75.041302}.
In general, the GCM describes the mixing between different particle-hole configurations, but we found that most of the excited states show weak mixing and are dominated by a single particle-hole configuration.
Therefore, we classify the excited states according to their dominant configurations.
We summarize the properties of the lowest energy state of each particle-hole configuration in Table \ref{table:31Mg}. We also show their quadrupole moments in the Appendix B.

\begin{table}[htb]
  \centering
  \caption{\label{table:31Mg} Properties of the low-lying states of $^{31}$Mg given in the ascending order of excitation energies. The columns represent spin-parities, particle-hole configurations, excitation energies (MeV), quadrupole deformation parameters $\beta$, and $\gamma$ (degrees), proton, neutron, and matter radii, matter diffuseness parameter (fm), and total reaction cross sections (mb) on a carbon target at 240 MeV/nucleon.}
\begin{ruledtabular}
\begin{tabular}{ccccccccccc}
$J^{\pi}$ & $m$p$n$h & $E_x$  & $\beta$ & $\gamma$ &$r_p$  & $r_n$  & $r_m$ & $a_m$ & $\sigma_R$\\
\hline
$1/2_{1}^{+}$ & 2p3h & 0.0 & 0.45 & 1  & 3.15 & 3.38 & 3.29 & 0.63 & 1349\\ 
$3/2_{1}^{-}$ & 3p4h & 0.66 & 0.50 & 12 & 3.18 & 3.44 & 3.34 & 0.66 & 1373 \\
$3/2_{2}^{-}$ & 1p2h & 0.89 & 0.37 & 0 & 3.09 & 3.29 & 3.21 & 0.59 & 1307 \\
$3/2_{2}^{+}$ & 0p1h & 1.35 & 0.17 & 1 & 3.06 & 3.23 & 3.17 & 0.53 & 1280 \\ 
\end{tabular}
\end{ruledtabular}
\end{table}

Experimentally, it is known that the ground state is the $1/2^+$ state with 2p3h configuration~\cite{nishibata_2019}, and the excited states with different particle-hole configurations appear above it. The present result reproduces the spin-parity of the ground state, and as discussed in Ref.~\cite{nishibata_2019}, the calculated $3/2_{1}^{-}$ (3p4h), $7/2_{1}^{-}$ (1p2h), and $3/2_{2}^{+}$ (0p1h) states are likely corresponding to the observed $3/2^{-}$, $7/2^{-}$ and $3/2^{+}$ states at 0.22 MeV, 0.46 MeV, and 0.67 MeV, respectively.

\begin{figure}[h!]
  \begin{center}
  \includegraphics[width=\columnwidth]{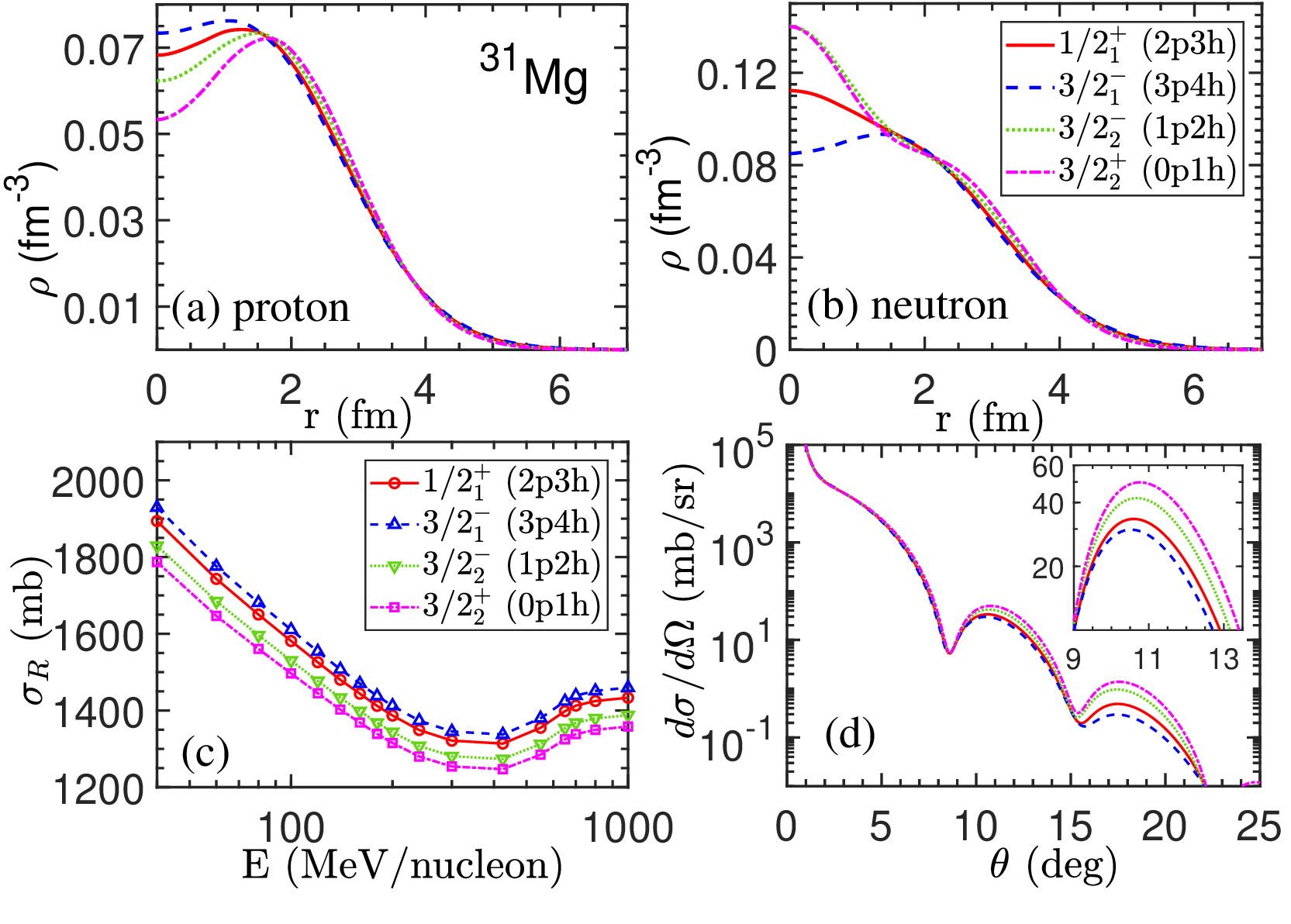}
  \caption{(a) Proton and (b) neutron density distributions of $^{31}$Mg, (c) total reaction cross sections on a carbon target at incident energies from 40 to 1000 MeV/nucleon, and (d) angular distribution of $^{31}\rm Mg$-proton elastic scattering cross sections at 800 MeV/nucleon for the states shown in Table~\ref{table:31Mg}. }
  \label{fig:dp-cs_31Mg}
  \end{center}
\end{figure}

Now, we examine the relationship between the particle-hole configuration and the nuclear surface diffuseness, which is the main focus of this study. 
In Figs.~\ref{fig:dp-cs_31Mg}~(a) and (b), we see that the central density of neutrons decreases with increasing the number of particles (holes) in the $pf$ ($sd$) shell, 
while the proton density shows the opposite trend. This is attributed to the variation in the occupation number of the $s$-wave, which changes with particle-hole excitation and deformation. Let us explain this for the case of the neutron densities (Fig.~\ref{fig:dp-cs_31Mg} (b)). 
Relative to the $3/2_2^{+}$ (0p1h) state, the $3/2_2^-$ (1p2h) state has a hole in the [2, 0, 2, 3/2] orbit and a particle in the [3, 3, 0, 1/2] orbit. Since neither of these orbits has $s$-wave mixing, the $3/2_2^+$ and $3/2_2^-$ states have almost equal occupation of the $s$-wave. Consequently, their central densities are nearly equal. On the other hand, the $1/2_1^+$ (2p3h) state has a hole in the [2, 0, 0, 1/2] orbit and a particle in the [3, 3, 0, 1/2] orbit on top of the $3/2_2^-$ (1p2h) state. As the hole in the [2, 0, 0, 1/2] orbit has $s$-wave mixing, the $1/2_1^+$ state has a reduced $s$-wave occupation, which suppresses the central density. The $3/2_1^-$ (3p4h) state has an additional hole in the [2, {\color{blue}0}, 0, 1/2] orbit, resulting in the most reduced central density. The behavior of the proton density can also be explained in a similar manner.

As discussed in Ref.~\cite{whoriuchi-ptep}, the surface diffuseness of the matter density distribution also depends on the single-particle configurations but in a different manner. In Tab.~\ref{table:31Mg}, we list the diffuseness parameters $a_m$ of the matter distribution, which are extracted by fitting the matter density distribution with the two-parameter Fermi distribution~\cite{hatakeyama_PhysRevC.97.054607};
\begin{align}
  \rho_m(r) = \frac{\rho_0}{1+\exp\left\{(r-R_m)/a_m\right\}},
\end{align}
where $R_m$ and $a_m$ are the radius and diffuseness parameters, respectively. The central-density parameter $\rho_0$ is determined by the normalization condition $4\pi\int_0^\infty \rho(r) r^2dr=A$.

Note that the nuclear surface diffuseness is roughly determined by the characteristics of the outermost orbits; in other words, it reflects the tail part of the weakly bound orbits~\cite{whoriuchi-ptep}. Orbits with smaller binding energy with lower orbital angular momentum, and larger nodes lead to a longer tail of the single particle wave function, and hence larger nuclear surface diffuseness. With this in mind, the variation of the diffuseness can be explained as follows: In the $3/2_2^{+}$ (0p1h) state, the valence neutron occupies the [2, 0, 2, 3/2] orbit, which is dominated by $d$-wave, resulting in a sharper density distribution with a diffuseness parameter $a_m = 0.53$. In the $3/2_2^{-}$ (1p2h) state, the valence neutron occupies the weakly bound [3, 3, 0, 1/2] orbit. Since this orbit has a mixture of the $p$-wave coming from the nodal $1p_{3/2}$ orbit, the diffuseness is increased. The diffuseness further increases in the $1/2_1^+$ (2p3h) and $3/2_1^-$ (3p4h) states by the occupation of the weakly bound [2, 0, 0, 1/2] and [3, 2, 1, 3/2] orbits with $s$-wave (from the $1s_{1/2}$ orbit) and $p$-wave mixing, respectively.  

Thus, the particle-hole configuration shows a reasonable correlation with the density distribution. It should be noted that the diffuseness of the nuclear surface density can be observed from the first peak of the elastic scattering cross section. In addition, an increase of particles and holes leads to the growth of the deformation and radius, leading to the enhancement of the total reaction cross section. Therefore, the elastic scattering and total reaction cross sections can be useful observables for estimating the type of the particle-hole configuration. 
 
To demonstrate this idea numerically, we calculate the total reaction cross section and angular distribution of the elastic scattering cross sections from the density distributions of different particle-hole configurations using the Glauber model.
Figure~\ref{fig:dp-cs_31Mg} (c) demonstrates that the total reaction cross section increases with increasing the matter radius, which is attributed to the enhancement of deformation. 
Note that the measured total reaction cross section at 240 MeV/nucleon is 1329(11) mb~\cite{Takechi_2014}, closely matching the calculated one for the $1/2^+_1$ state with 2p3h configuration, thereby reconfirming it as the ground state of $^{31}{\rm Mg}$.

Different information can be obtained from the elastic scattering cross sections. 
As explained in Ref.~\cite{hatakeyama_PhysRevC.97.054607}, if the diffuseness parameter is large, incident particles scattered at the nuclear surface slightly deviate in the scattering angles (in a classical sense). As a result, the diffraction is weakened, that is, the peak height decreases. 
The calculated results shown in Fig.~\ref{fig:dp-cs_31Mg} (d) precisely exhibit this trend.
The first diffraction peak height of the $3/2^-_1$ (3p4h) state with the largest diffuseness is 30 mb/sr, which is approximately 60\% smaller than the $3/2^+_2$(0p1h) state.
Thus, the particle-hole configurations affect the cross sections, and by using this correlation, we could identify the spin-parity of unknown nuclei.

\subsection{Cross sections for $^{29}\rm Ne$, $^{33}\rm Mg$ and $^{35}\rm Mg$}\label{subsection:IIIB}
In this section, we will quantitatively examine the correlation between particle-hole configurations and cross sections for neighboring nuclei and discuss to what extent the cross sections can be used to identify/exclude the spin-parity assignments.
 
\subsubsection{$^{29}\rm Ne$} 
The spin-parity of $^{29}$Ne remains uncertain. The measurements of its $\beta$-decay~\cite{vtripathi_PhysRevLett.94.162501} reported a large branching ratio to the 5/2$^+$ ground state of $^{29}$Na, favoring 3/2$^+$ assignment to $^{29}$Ne. 
In contrast, the one-neutron removal experiment~\cite{kobayashi-prc2016} suggested 3/2$^-$ assignment based on the observed parallel momentum distribution. 
Another study also measuring the one-neutron removal cross sections~\cite{LIU201758} had insufficient energy resolution for low-lying states, detecting both $d$- and $p$-wave components, and hence implies the coexistence of the 3/2$^-$ and 3/2$^+$ states below 200 keV. 

On the theory side, a shell model calculation with a standard interaction SDPF-M~\cite{LIU201758} predicted the 3/2$^+$ ground state with the 3/2$^-$ state 70 keV above. 
However, it was also shown that changes in the interaction and model space can invert this order~\cite{LIU201758}. Hence, the spin-parity of this nucleus could be a good test ground for effective interactions.

\begin{figure}[tb!]
  \begin{center}
  \includegraphics[width=0.95\hsize]{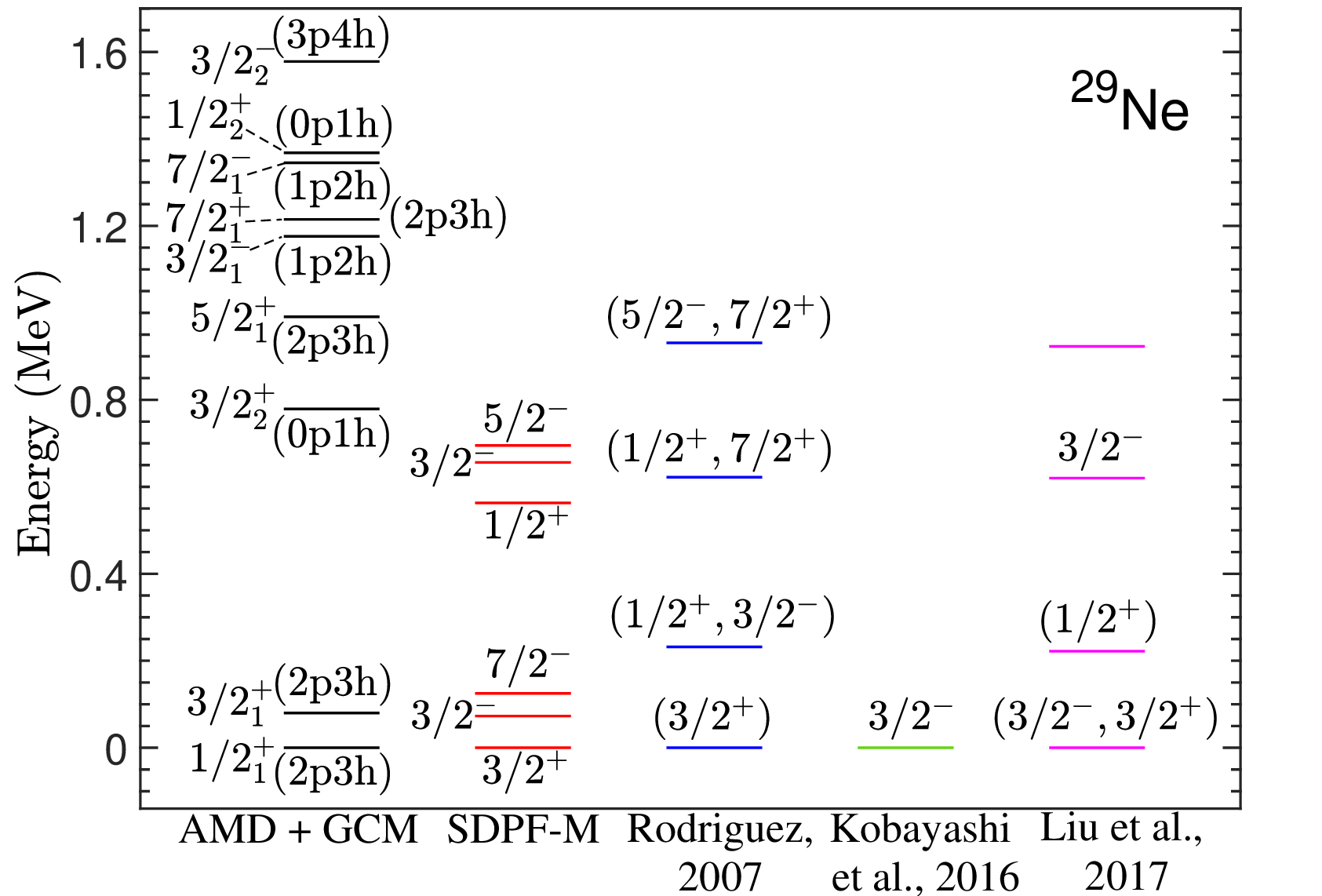}
  \caption{Excitation spectra of  $^{29}$Ne. The shell model results and experimental data are taken from Refs.~\cite{kobayashi-prc2016, rodriguez_2007, LIU201758}}
  \label{fig:ex29Ne}
  \end{center}
\end{figure}
The energy spectrum obtained by the present calculation with the Gogny D1S interaction is shown in Fig.~\ref{fig:ex29Ne} together with the shell model calculations as well as the experimental data. As $^{29}$Ne has the same neutron number $N=19$ with $^{31}\rm Mg$, the dominant particle-hole configurations for the low-lying states are the same. The present calculation suggests the 1/2$^+$ (2p3h) ground state with the 3/2$^+$ (2p3h) state 100 keV above, and the 3/2$^-$ (3p4h) state at 1.58 MeV, which quantitatively differs from the shell model predictions. However, as in the shell model calculations, the level ordering is sensitive to the interaction used, \textit{e.g.}, employing the Gogny D1M interaction changes the ground state to $3/2^+$ and shifts the 3/2$^-$ state down to 200 keV.

\begin{figure}[h!]
    \begin{center}
    \includegraphics[width=\columnwidth]{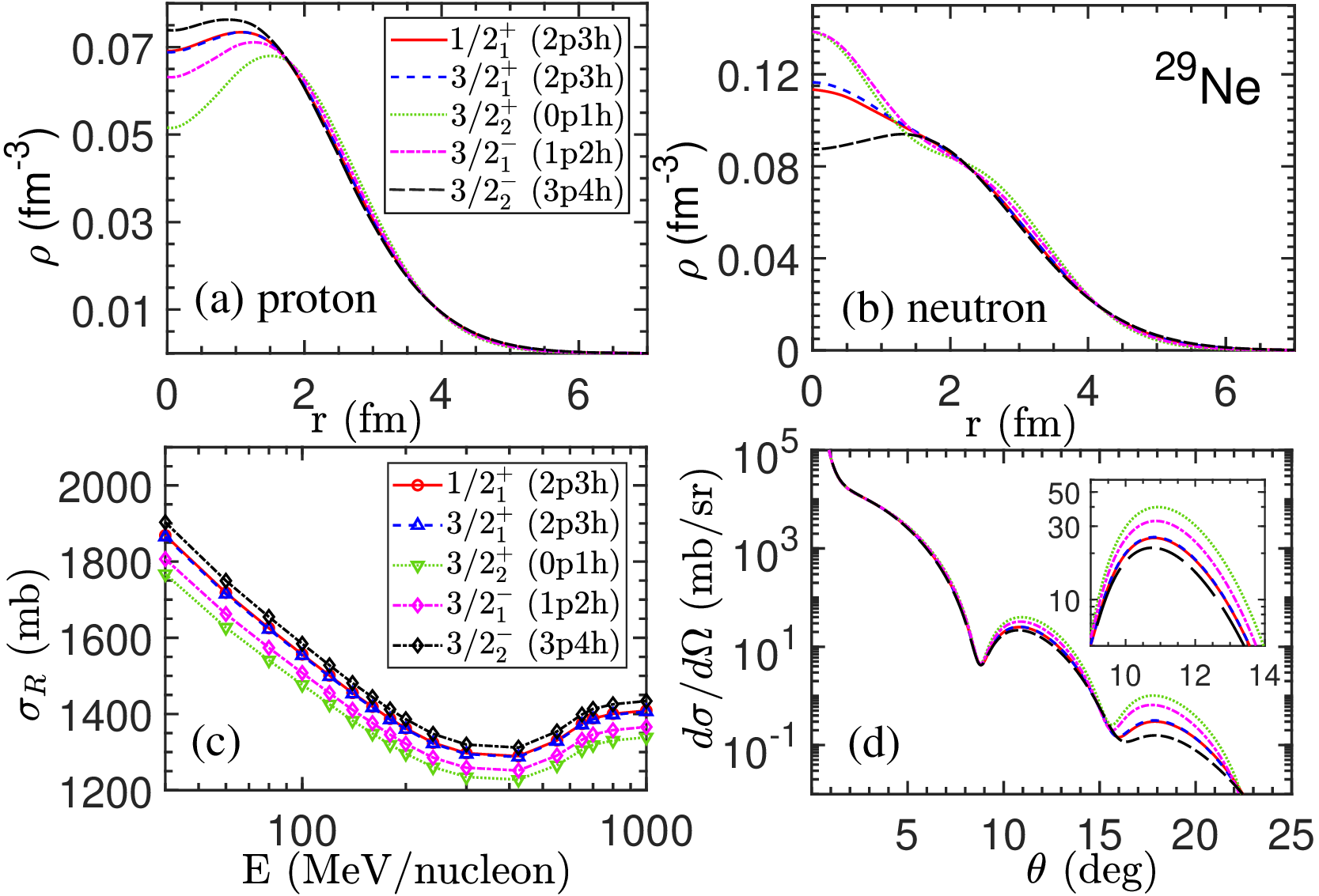}
    \caption{Same as Fig. \ref{fig:dp-cs_31Mg} but for $^{29}\rm Ne$.}
    \label{fig:dp-cs_29Ne}
    \end{center}
\end{figure}

As candidates of the ground state, Table \ref{table:29Ne} lists the $3/2^{\pm}$ states obtained with the Gogny D1S interaction in addition to the $1/2^+$ state. Their proton and neutron density distributions are shown in Fig. \ref{fig:dp-cs_29Ne}~(a) and (b). Due to the similarity in the single-particle orbits, $^{29}\rm Ne$ shows the behavior of the density profiles (central density and diffuseness) analogous to $^{31}\rm Mg$.
\begin{table}[h!]
  \centering
  \caption{\label{table:29Ne}Properties of low-lying states of $^{29}$Ne. The table columns are represented in the same way as described in Table \ref{table:31Mg}.}
\begin{ruledtabular}
\begin{tabular}{ccccccccccc}
$J^{\pi}$ & $m$p$n$h & $E_x$  & $\beta$ & $\gamma$ & $r_p$  & $r_n$  & $r_m$ & $a_m$ &$\sigma_R$\\
\hline
$1/2_{1}^{+}$ & 2p3h & 0.0 & 0.45 & 0  & 3.05 & 3.38 & 3.27 & 0.65& 1325\\ 
$3/2_{1}^{+}$ & 2p3h & 0.08 & 0.42 & 0 & 3.04 & 3.38 & 3.27 & 0.65& 1322\\
$3/2_{2}^{+}$ & 0p1h & 0.78 & 0.22 & 33 &  2.98 & 3.25 & 3.16 & 0.54& 1261\\ 
$3/2_{1}^{-}$ & 1p2h & 1.18 & 0.30 & 0 & 3.00 & 3.30 & 3.20 & 0.59& 1285\\
$3/2_{2}^{-}$ & 3p4h & 1.58 & 0.52 & 0 & 3.07 & 3.44 & 3.32 & 0.68 & 1348\\
\end{tabular}
\end{ruledtabular}
\end{table}
\begin{table}[h!]
  \centering
  \caption{\label{table:29Ne-D1M}Same as Table \ref{table:29Ne}, but the calculations are done using Gogny D1M interaction.}
\begin{ruledtabular}
\begin{tabular}{ccccccccccc}
$J^{\pi}$ & $m$p$n$h & $E_x$  & $\beta$ & $\gamma$ & $r_p$  & $r_n$  & $r_m$ & $a_m$ &$\sigma_R$\\
\hline
$3/2_{1}^{+}$ & 2p3h+0p1h & 0.0 & 0.45 & 0  & 2.98 & 3.27 & 3.17 & 0.59& 1276\\ 
$1/2_{1}^{+}$ & 2p3h & 0.1 & 0.45 & 0 & 2.99 & 3.30 & 3.19 & 0.60& 1286\\
$3/2_{1}^{-}$ & 3p4h & 0.2 & 0.53 & 10 &  3.04 & 3.40 & 3.28 & 0.67& 1331\\ 
$3/2_{2}^{-}$ & 1p2h & 0.9 & 0.32 & 4 & 2.96 & 3.25 & 3.16 & 0.58& 1267\\
$3/2_{2}^{+}$ & 0p1h+2p3h & 1.23 & 0.2 & 24 & 2.96 & 3.25 & 3.15 & 0.57 & 1265\\
\end{tabular}
\end{ruledtabular}
\end{table}

The calculated total reaction cross sections are shown in Fig. \ref{fig:dp-cs_29Ne}(c). The observed interaction cross section $\sigma_I$ at 240 MeV/nucleon is 1344(13) mb~\cite{Takechi_2010},which is comparable to the calculated $\sigma_R$ for the $3/2_2^-$ (3p4h) state, 1348 mb,\footnote{For the weakly bound nuclei, which has a few bound excited states, $\sigma_I$ and $\sigma_R$ are considered to be approximately equal in such high-energy reactions~\cite{horiuchi_PhysRevC.86.024614,horiuchi_aris2014}.}, indicating that the $3/2^-$ state with a 3p4h configuration is likely the ground state of $^{29}\rm{Ne}$ as proposed in Ref.~\cite{kobayashi-prc2016}. 

We also note that $^{29}\rm Ne$ is weakly bound and its one-neutron separation energy depends on the choice of effective interaction, which may influence the radial wave functions and slightly change the cross sections. To verify this uncertainty,  we examine the result with Gogny D1M interaction and present them in Table \ref{table:29Ne-D1M}. As mentioned before, Gogny D1M results give deeper binding energy for the $3/2^-$ (3p4h) state, which, in effect, gives a lesser spatial extension of
the wave function. We indeed found a slight reduction in both its radius and diffuseness, which also results in a $1.2\%$ decrease in $\sigma_R$ compared to the Gogny D1S case. We further find that changing the interaction affects the degree of configuration mixing in some cases. For instance, the $3/2_1^+$ and $3/2_2^+$ states have an enhanced mixing of 0p1h and 2p3h configurations, leading to a reduction in the radii and diffuseness of the $3/2_1^+$ state compared to the Gogny D1S case. Consequently, only the $3/2^-$ state with 3p4h configuration can explain the observed cross section \cite{Takechi_2010}. This assignment,  the $3/2^-$ (3p4h) state as the ground state, is consistent with the experiment \cite{kobayashi-prc2016}, and is independent of the choice of interaction. 

Now let us see the angular distribution of the elastic scattering cross sections calculated using the Gogny D1S interaction. As shown in the inset of Fig.~\ref{fig:dp-cs_29Ne}~(d),
the states are distinguishable from differences in their cross sections at the first diffraction peak, except for the $1/2_1^{+}$ (2p3h) and $3/2_1^{+}$ (2p3h) states, which are members of the same $K^\pi=1/2^+$ rotational band and have a similar diffuseness parameter.
Thus, the measurement of the elastic scattering cross sections may give us an additional insight into the spin-parity of $^{29}$Ne.
For the Gogny D1M case, the differences in cross sections at the first peak for $1/2^+$ and $3/2^+$ states are slightly reduced compared to the Gogny D1S case as their diffuseness parameters are not very different due to the enhancement of configuration mixing. However, the $3/2_1^-$ (3p4h) state still exhibits a significant, and even more pronounced difference from the other states.

\subsubsection{$^{33}\rm Mg$}
\begin{figure}[h!]
    \begin{center}
    \includegraphics[width=\columnwidth]{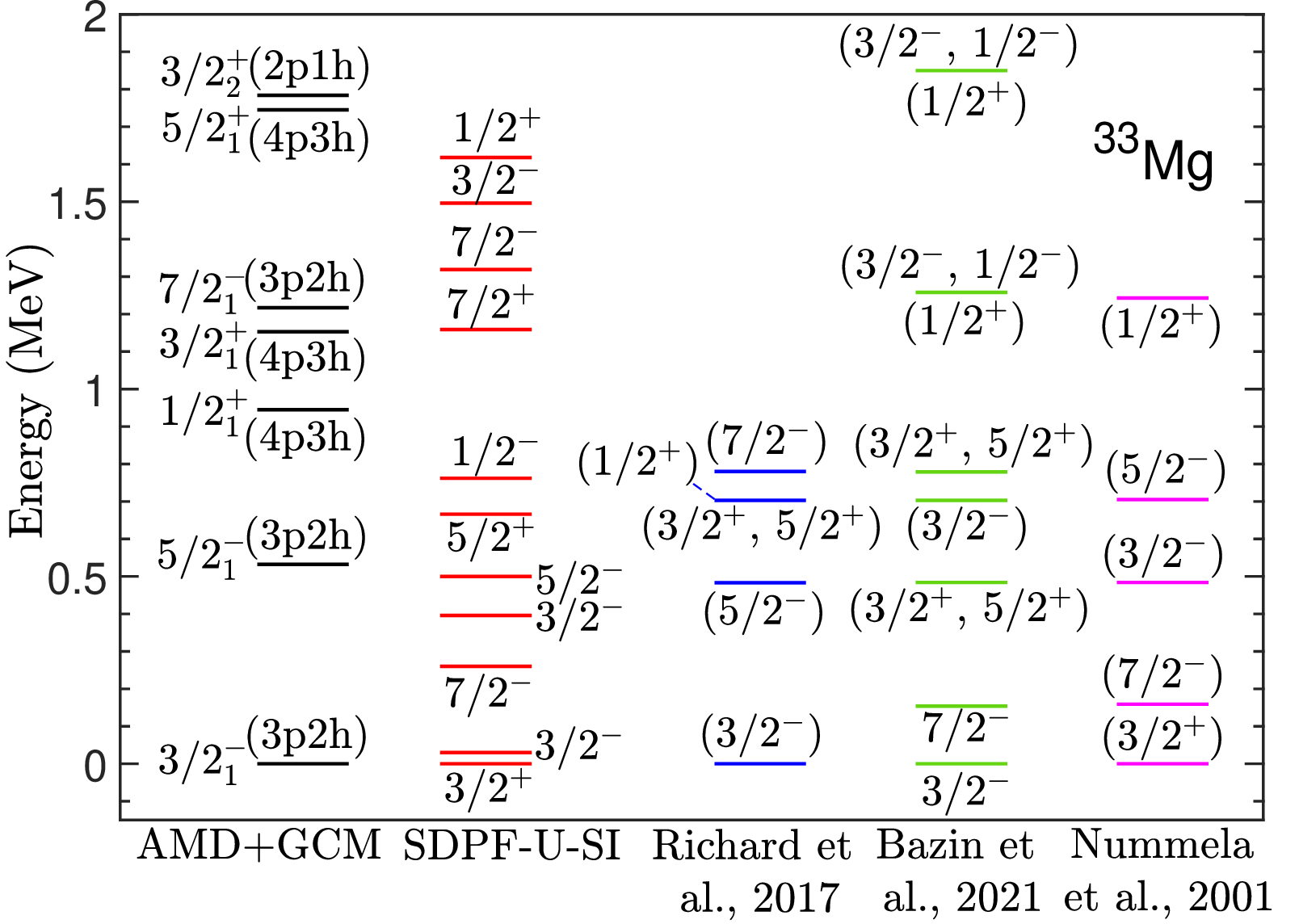}
    \caption{Excitation spectra of  $^{33}$Mg. The shell model results and experimental data are taken from Refs.~\cite{richard_2017, bazin2021, nummela_2001}.}
    \label{fig:ex33Mg}
    \end{center}
\end{figure}

To discuss $^{33}\rm Mg$, we start with the energy spectra shown in Fig. \ref{fig:ex33Mg}. Recent studies \cite{richard_2017, bazin2021} reported that the ground state of $^{33}\rm Mg$ is likely to be $3/2^-$. However, in the past, discussions of a $3/2^{+}$ ground state were made based on the $\beta$-decay measurements from $^{33}\rm Na$ \cite{nummela_2001}. Many theoretical calculations \cite{bazin2021}, including our present works, predict a $3/2^-$ ground state, while the shell model calculation with the SDPF-U-SI interaction predicts a $3/2^+$ ground state. To test resolve the issue in these contradicting spin-parity assignments, we investigate the low-lying $3/2^{\pm}$ states with different particle-hole configurations, whose properties are listed in Tab.~\ref{table:33Mg}. 

\begin{table}[h!]
    \centering
    \caption{\label{table:33Mg}Properties of low-lying states of $^{33}$Mg. The table columns are represented in the same way as Table \ref{table:31Mg} and \ref{table:29Ne}. }

\begin{ruledtabular}
\begin{tabular}{ccccccccccc}

 $J^{\pi}$ & $m$p$n$h & $E_x$ & $\beta$ & $\gamma$ & $r_p$  & $r_n$  & $r_m$ & $a_m$ & $\sigma_R$\\
 \hline
 
$3/2_{1}^{-}$ & 3p2h & 0.0 & 0.40 & 0 & 3.17 & 3.46 & 3.36 & 0.64 &1393\\ 
$3/2_{1}^{+}$ & 4p3h & 1.15 & 0.52 & 14 & 3.22 & 3.52 & 3.41 & 0.68 &1423\\
$3/2_{2}^{+}$ & 2p1h & 1.78 & 0.32 & 0 & 3.15 & 3.39 & 3.31 & 0.60 &1365\\ 
\end{tabular}
\end{ruledtabular}
\end{table}

\begin{figure}[h!]
    \begin{center}
    \includegraphics[width=\columnwidth]{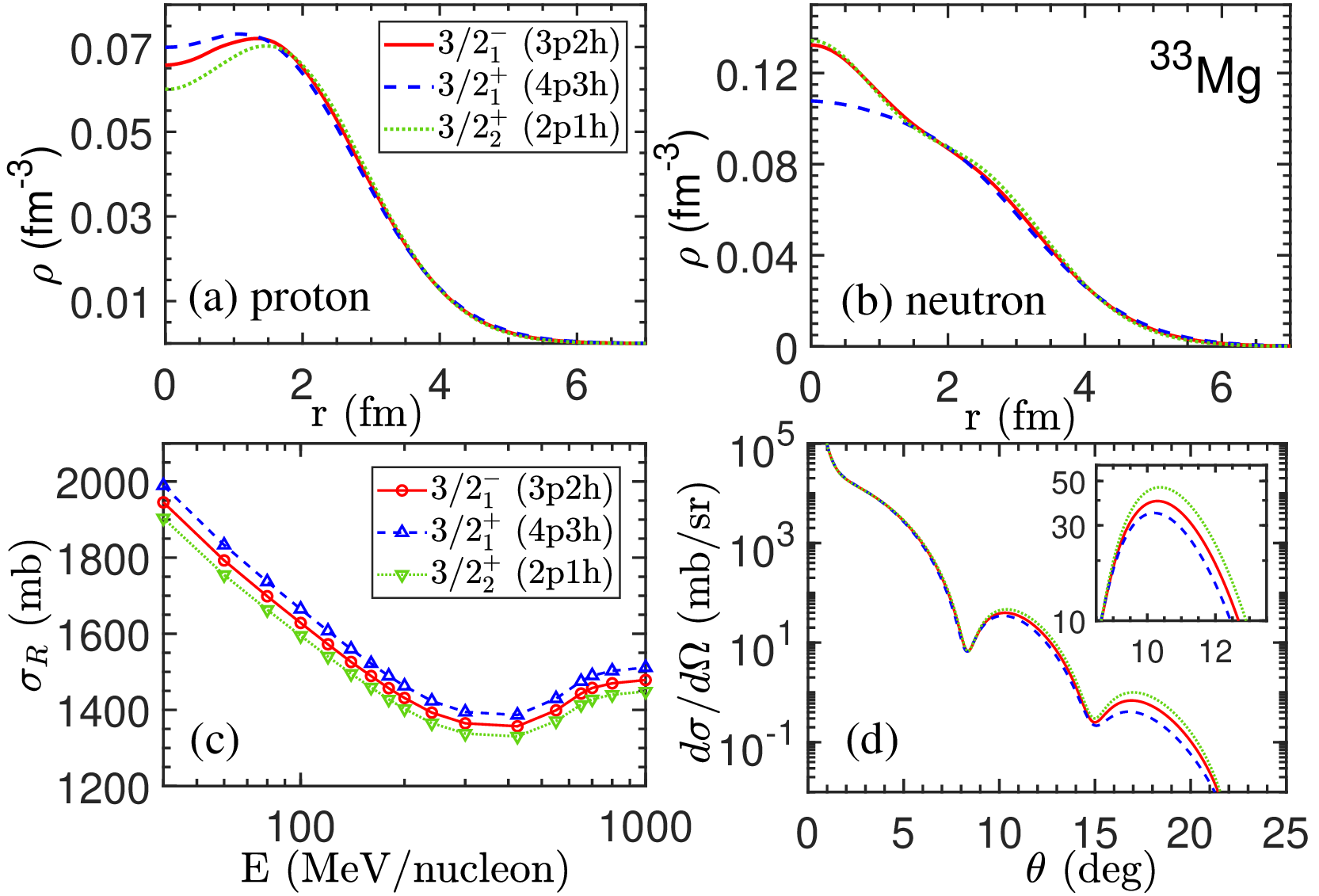}
    \caption{Same as Fig. \ref{fig:dp-cs_31Mg} but for $^{33}\rm Mg$.}
    \label{fig:dp-cs_33Mg}
    \end{center}
\end{figure} 
The proton and neutron density distributions are shown in Fig. \ref{fig:dp-cs_33Mg}(a) and (b). The central densities vary depending on the $s$-wave occupancy, and the diffuseness changes with the energy and angular momentum of the valence neutron as discussed for $^{31}\rm Mg$.

Fig. \ref{fig:dp-cs_33Mg}~(c) and (d) displays the calculated cross sections. The measured total reaction cross section at 240 MeV/nucleon is 1399(12) mb~\cite{Takechi_2014}, which is closest to the calculated $\sigma_R$ for the $3/2^-$ (3p4h) state, 
1393 mb~\cite{Takechi_2014}. The other two $3/2^+$ states significantly deviate from the measured value, indicating that the $3/2^-$ state is the ground state of $^{33}\rm Mg$. The calculated quadrupole moment further supports this assignment (Appendix B).
The magnitudes of the first diffraction peak in the elastic scattering cross sections also distinguish each state, as shown in the inset of Fig. \ref{fig:dp-cs_33Mg}.

\subsubsection{$^{35}\rm Mg$}

For $^{35}\rm Mg$, the ground-state spin-parity has not been established yet. Ref.~\cite{gade2011} tentatively assigned a $5/2^-$ ground state based on the observed gamma-ray transitions. Another study in Ref. \cite{35mg_momiyama} reported a few low-lying states and showed significant mixing of the $p$- and $f$-wave components in the inclusive parallel momentum distributions of the one-neutron knockout reaction. Theoretical studies also show contradictory results. The shell model study using the SDPF-M and SDPF-M+$p_{1/2}$ interactions suggested a $3/2^-$ ground state. The $5/2^-$ or $1/2^-$ states lie very close to the ground state in these calculations. The $3/2^+$ state also appears at a small excitation energy of approximately 600--800 keV. 

\begin{figure}[h!]
  \begin{center}
  \includegraphics[width=\columnwidth]{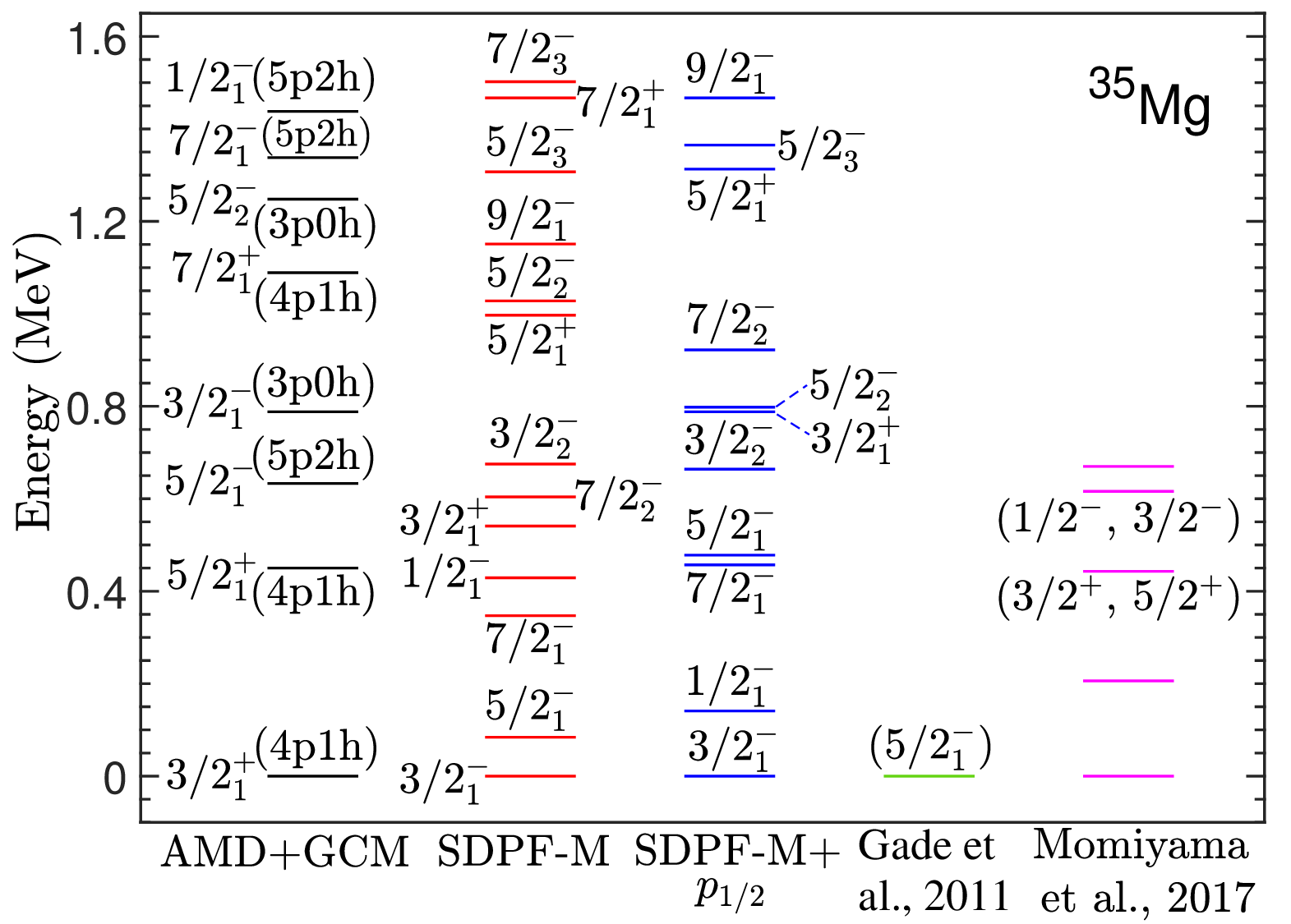}
  \caption{Excitation spectra of $^{35}$Mg. The shell model results and experimental data are taken from Refs.~\cite{35mg_momiyama, gade2011}.}
  \label{fig:ex35Mg}
  \end{center}
\end{figure}

\begin{table}[h!]
  \centering
  \caption{\label{table:35Mg} Properties of the low-lying states of $^{35}$Mg. The table columns are represented in the same way as Table \ref{table:31Mg}, \ref{table:29Ne}, and \ref{table:33Mg}. }

\begin{ruledtabular}
\begin{tabular}{cccccccccc}

$J^{\pi}$ & $m$p$n$h &  $E_x$  & $\beta$ & $\gamma$ & $r_p$  & $r_n$  & $r_m$ & $a_m$ & $\sigma_R$\\
\hline

$3/2_{1}^{+}$ & 4p1h & 0.0 & 0.40 & 0 & 3.19 & 3.51 & 3.40 & 0.63 & 1427\\ 
$5/2_{1}^{-}$ & 5p2h & 0.63 & 0.45 & 22 & 3.21 & 3.55 & 3.44& 0.66 & 1444\\
$3/2_{1}^{-}$ & 3p0h & 0.79 & 0.32 & 0 & 3.17 & 3.46 & 3.36 & 0.60 & 1403\\

\end{tabular}
\end{ruledtabular}
\end{table}

Our calculations predict the $3/2_1^{+}$ ground state with a 4p1h configuration together with the $3/2^-_1$ state at 800 keV. We list the properties of these potential ground states: $3/2_1^{+}$ (4p1h), $5/2_{1}^-$ (5p2h), and $3/2_1^{-}$ (3p0h) in Tab.~\ref{table:35Mg}, as they are the lowest energy states with different particle-hole configurations in our calculation.

\begin{figure}[h!]
    \begin{center}
    \includegraphics[width=\columnwidth]{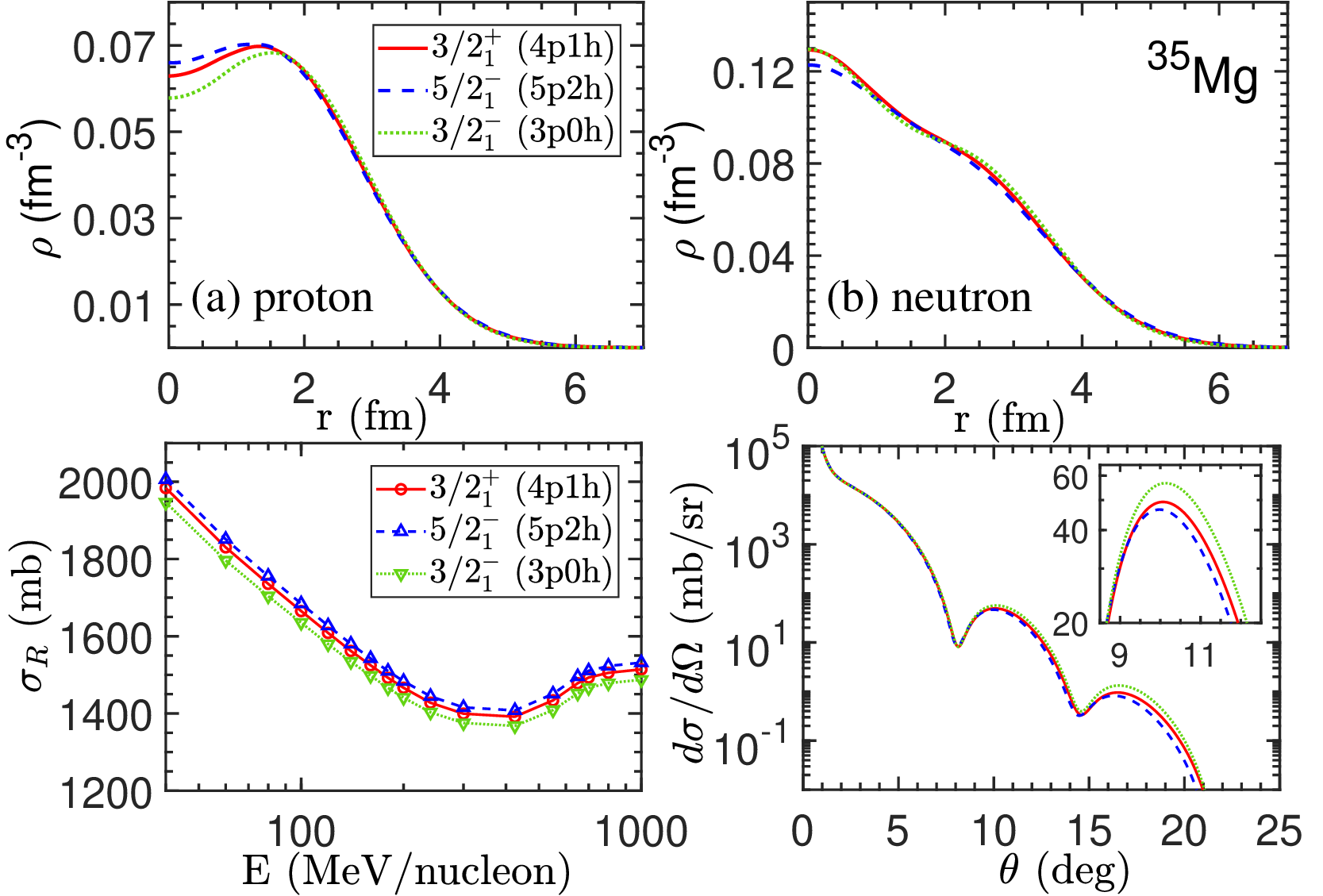}
    \caption{Same as Fig. \ref{fig:dp-cs_31Mg}, but for $^{35}\rm Mg$.}
    \label{fig:dp-cs_Mg35}
    \end{center}
\end{figure} 

The density distribution shown in Fig. \ref{fig:dp-cs_Mg35}~(a) and (b) exhibit less variation compared to lighter isotopes. This is likely due to a decrease in the number of holes and the increase of particles in the $pf$-shell. A particularly intriguing aspect is that the $5/2_1^-$ (5p2h) state shows a reduced central density despite having a fully occupied [2, 0, 0, 1/2] orbit, which is supposed to be dominated by the $s$-wave. This can be explained by the mixing of the $g$-wave components in this orbit induced by the large deformation.

The surface diffuseness is also less sensitive to the particle-hole configurations, as seen in Tab.~\ref{table:35Mg}. Since many neutrons always occupy the weakly bound $pf$-shell, the surface diffuseness does not change significantly with small changes in the number of particles and holes.

We show the total reaction cross sections in Fig. \ref{fig:dp-cs_Mg35}~(c). The experimental value of the total reaction cross section at 240 MeV/nucleon is 1443(12) mb~\cite{Takechi_2014}. Our calculation gives 1444 mb for the $5/2_1^{-}$ (5p2h) state, which is an excellent agreement. However, since the difference in the cross sections is not significant, we cannot strictly exclude the other spin-parity states based on this result alone.

Similarly, the angular distributions of the elastic scattering cross sections shown in Fig.~\ref{fig:dp-cs_Mg35}~(d) do not exhibit large differences in the height of the first diffraction peak.
The height of the first peak for the $3/2_1^+$(4p1h) state and the $5/2_1^-$ (5p2h) state differ by only 2.7 mb/sr, which is relatively smaller compared to those in the previously discussed nuclei. Given this situation, $^{35}\rm Mg$ may not be a good case 
for the spin-parity assignment by this method.

\section{Summary and conclusions}\label{sec:conclusion}
We have discussed the relationship between the particle-hole configurations, the density profiles and total reaction and elastic cross sections for the nuclei in the $N=20$ island of inversion. 

First, the $^{31}$Mg nucleus, whose low-lying spectrum is well understood, is studied as a test case to find the correlation between the particle-hole configurations and density profiles.
We have shown that both the central density and surface diffuseness of the density distribution are sensitive to the particle-hole configurations. Furthermore, Glauber model calculations revealed that the total reaction cross section and the angular distribution of the elastic scattering cross sections are sensitive to the density profile and consequently to the particle-hole configurations.

Encouraged by these results, we have explored the possibility of determining the spin-parities of the controversial $^{29}$Ne, $^{33}$Mg, and $^{35}$Mg nuclei from these cross sections. 
Earlier studies suggested two possible ground-state spin-parities for $^{33}$Mg: $3/2^-$ and $3/2^+$. The calculated total reaction cross section for the $3/2^-$ state shows an excellent agreement with the experimental data, while the $3/2^+$ state does not. Thus, we reconfirm the $3/2^-$ ground state for $^{33}$Mg. 
Additionally, the height of the first diffraction peak in the elastic scattering cross sections clearly distinguishes them, which will serve as additional proof.
In the case of $^{29}$Ne, the calculated $3/2^-_2$ (3p4h) state with Gogny D1S interaction exhibited better agreement with the measured total interaction cross section compared to the $1/2^+_1$ state, which was predicted as the ground state by the AMD calculations. However, when the interaction is changed to Gogny D1M, the ground state becomes $3/2^+$, and the total reaction cross section for $3/2^-$ (3p4h) slightly change due to deeper binding energy. Nevertheless, it still remains consistent with the experimental value for both interactions.
This suggests that the $3/2^-$ assignment proposed by Kobayashi et al.~\cite{kobayashi-prc2016} might be correct and may indicate the need for the refinement of the effective interaction in the theoretical calculation. The peak height of the first diffraction of the elastic scattering cross sections also differs in the  $3/2^-$ (3p4h) and the other states in both interactions, which can further clarify the assignment of this spin-parity if measured.
Finally, for $^{35}$Mg, the differences in the cross sections between different particle-hole configurations are not as significant compared to the other cases. This is due to the larger number of neutrons occupying the $pf$-shell, which reduces the impact of the holes in the $sd$-shell on the density distribution. Therefore, having the cross sections alone is not enough conclusive to determine the spin-parity of $^{35}$Mg.

To assess the experimental feasibility of distinguishing different configurations from the first peak of the elastic scattering cross-section, we examine the available data for the stable isotopes of these nuclei. The data shows cross sections in the same order as in our calculations with significantly small error bars \cite{blanpied_PhysRevC.38.2180, Blanpied_PhysRevC.37.1987}. In this context, if similar measurements for the discussed exotic nuclei yield cross sections of the same order, a clear distinction between these states can be expected.

We expect that the method proposed here can be applied to other mass regions as well. For example, it would be interesting to apply it to the $N=28$ island of inversion, where a variety of nuclear deformation is expected.

\acknowledgements
This work was supported by JSPS Bilateral Program Number JPJSBP120247715 and DST/INT/JSPS/P-393/2024(G).  The work was in part supported by JSPS KAKENHI Grants Nos. 23K22485, 25K07285, and 25K01005. R.B. acknowledges RIKEN for the International Program Associate (IPA) fellowship.

\section*{Appendix}\label{appendix}
\subsection{Definition of $\beta$ and $\gamma$}\label{deformation}

The quadrupole deformation pararmeters $\beta$ and $\gamma$ are defined following the prescription given in \cite{kimura_10.1143/PTP.127.287}, which is given as follows:
\begin{align}
    \left<x^2\right> &= R_0^2\left[1+\sqrt{\frac{5}{\pi}}\beta \cos(\gamma+2\pi/3)\right],\\
    \left<y^2\right> &= R_0^2\left[1+\sqrt{\frac{5}{\pi}}\beta \cos(\gamma-2\pi/3)\right],\\
    \left<z^2\right> &= R_0^2\left[1+\sqrt{\frac{5}{\pi}}\beta \cos{\gamma} \right],\\
    R_0^2 &= \frac{1}{3}\left[\left<x^2\right>+ \left<y^2\right> + \left<z^2\right>\right],
\end{align}

where $\left<x^2\right>$, $\left<y^2\right>$, and $\left<z^2\right>$ represent the expectation values of the respective operators calculated using the intrinsic wave function. The intrinsic coordinate system is defined such that the condition $\left<x^2\right> \leq \left<y^2\right> \leq \left<z^2\right>$ is satisfied.

\subsection{Quadrupole moments}\label{qmoment}

We tabulate the calculated electric quadrupole moments for the different low-lying states for all the nuclei discussed in this paper. It is a good observable to validate the reliability of our calculations and the deformation parameters. However, the experimental data is not available for most of the nuclei. For $^{33}$Mg, the experimental $Q$ for the ground state is 13(9) $\rm e^2fm^4$, to which our calculation gives 14.82 $\rm{e^2fm^4}$, which is in well agreement with the experimental value.

\begin{table}[h!]
  \centering
  \caption{Electric quadrupole moments for $^{31}\rm Mg$.}

\begin{ruledtabular}
\begin{tabular}{c|cccc}

$J^\pi$ & $1/2_1^+$ & $3/2_1^-$ & $3/2_2^-$ & $3/2_2^+$\\
\hline\\
$Q$ ($\rm e^2fm^4$) &-&17.1&-10.9&6.8\\

\end{tabular}
\end{ruledtabular}
\end{table}

\begin{table}[h!]
  \centering
  \caption{Electric quadrupole moments for $^{29}$Ne.}

\begin{ruledtabular}
\begin{tabular}{c|ccccc}

$J^\pi$ & $1/2_1^+$ & $3/2_1^+$ & $3/2_2^+$ & $3/2_1^-$ & $3/2_2^-$\\
\hline\\
$Q$ ($\rm e^2fm^4$) &-&-11.49&6.42&-8.75& 13.26\\
\end{tabular}
\end{ruledtabular}
\end{table}

\begin{table}[h!]
  \centering
  \caption{Electric quadrupole moments for $^{33}\rm Mg$.}

\begin{ruledtabular}
\begin{tabular}{c|cccc}

$J^\pi$ & $3/2_1^-$ & $3/2_1^+$ & $3/2_2^+$\\
\hline\\
$Q$ ($\rm e^2fm^4$) &14.82 [13(9)]&-17.60&-12.57\\

\end{tabular}
\end{ruledtabular}
\end{table}

\begin{table}[h!]
  \centering
  \caption{Electric quadrupole moments for $^{35}\rm Mg$.}

\begin{ruledtabular}
\begin{tabular}{c|cccc}

$J^\pi$ & $3/2_1^+$ & $5/2_1^-$ & $3/2_1^-$\\
\hline\\
$Q$ ($\rm e^2fm^4$) &14.53&-5.35&-11.96\\

\end{tabular}
\end{ruledtabular}
\end{table}

\bibliography{references}

\end{document}